\documentclass[11pt]{article}
\usepackage{blois,epsfig}
\usepackage[latin1]{inputenc}
\usepackage[OT1]{fontenc}

\def\Journal#1#2#3#4{{#1} {\bf #2}, #3 (#4)}

\def\RNC{\em Rivista Nuovo Cimento}

\def\NIMA{{\em Nucl. Instrum. Methods} A}

\def\PLB{{\em Phys. Lett.}  B}
\def\PRL{\em Phys. Rev. Lett.}
\def\PRD{{\em Phys. Rev.} D}

\def\GaC{\em Gravitation and Cosmology}
\def\GaCS{{\em Gravitation and Cosmology} Supplement}

\def\JETPL{\em JETP Lett.}
\def\PAN{\em Phys.Atom.Nucl.}
\def\CQG{\em Class. Quantum Grav.}
\def\APJ{\em Astrophys. J.}
\def\SCI{\em Science}

\def\mco{\multicolumn}

\def\ra{\rightarrow}

\def\ko{K^0}

\def\s{{\,\rm s}}
\def\g{{\,\rm g}}
\def\eV{\,{\rm eV}}
\def\keV{\,{\rm keV}}
\def\MeV{\,{\rm MeV}}

\def\TeV{\,{\rm TeV}}
\def\sv{\left<\sigma v\right>}
\def\({\left(}
\def\){\right)}
\def\cm{{\,\rm cm}}

\def\kpc{{\,\rm kpc}}
\def\beq{\begin{equation}}
\def\eeq{\end{equation}}
\def\bea{\begin{eqnarray}}
\def\eea{\end{eqnarray}}

\begin{document}
\vspace*{4cm}
\title{COMPOSITE DARK MATTER FROM STABLE CHARGED CONSTITUENTS}

\author{ M.Y. KHLOPOV }

\address{Moscow Engineering Physics Institute, Moscow, Russia;\\
Center for Cosmoparticle physics ``Cosmion'', Moscow, Russia;\\
APC laboratory 10, rue Alice Domon et L\'eonie Duquet 75205 Paris
Cedex 13, France}

\maketitle\abstracts{ Heavy stable charged particles can exist,
hidden from us in bound atomlike states. Models with new stable
charged leptons and quarks give rise to realistic composite dark
matter scenarios. Significant or even dominant component of O-helium
(atomlike system of He4 nucleus and heavy -2 charged particle) is
inevitable feature of such scenarios. Possible O-helium explanation
for the positron excess in the galactic bulge and for the
controversy between the positive results of DAMA and negative
results of other experiments is proposed.}

\section{Introduction}
The widely shared belief is that the dark matter, corresponding to
$25\%$ of the total cosmological density, is nonbaryonic and
consists of new stable particles. One can formulate the set of
conditions under which new particles can be considered as candidates
to dark matter (see e.g. \cite{book,Cosmoarcheology,Bled07} for
review and reference): they should be stable, saturate the measured
dark matter density and decouple from plasma and radiation at least
before the beginning of matter dominated stage. The easiest way to
satisfy these conditions is to involve neutral weakly interacting
particles. However it is not the only particle physics solution for
the dark matter problem. As we show here, new stable particles can
have electric charge, but escape experimental discovery, because
they are hidden in atom-like states maintaining dark matter of the
modern Universe.

Recently several elementary particle frames for heavy stable charged
particles were proposed:  (a)  A heavy quark of fourth generation
\cite{I,lom,Khlopov:2006dk} accompanied by heavy neutrino \cite{N};
which can avoid experimental constraints \cite{Q,Okun} and form
composite dark matter species; (b) A Glashow's ``sinister'' heavy
tera-quark $U$ and tera-electron $E$, forming a tower of
tera-hadronic and tera-atomic bound states with ``tera-helium
atoms'' $(UUUEE)$ considered as dominant dark matter
\cite{Glashow,Fargion:2005xz}. (c) AC-leptons, predicted in the
extension \cite{5} of standard model, based on the approach of
almost-commutative geometry \cite{bookAC}, can form evanescent
AC-atoms, playing the role of dark matter
\cite{5,FKS,Khlopov:2006dk}. Finally, it was recently shown in
\cite{KK} that an elegant composite dark matter solution is possible
in the framework of walking technicolor models (WTC)
\cite{Sannino:2004qp}.

In all these models (see review in
\cite{Khlopov:2006dk,Bled07,Khlopov:2008rp}), the predicted stable
charged particles form neutral atom-like states, composing the dark
matter of the modern Universe. It offers new solutions for the
physical nature of the cosmological dark matter. The main problem
for these solutions is to suppress the abundance of positively
charged species bound with ordinary electrons, which behave as
anomalous isotopes of hydrogen or helium. This problem is
unresolvable, if the model predicts stable particles with charge -1,
as it is the case for tera-electrons \cite{Glashow,Fargion:2005xz}.

The possibility of stable doubly charged particles $A^{--}$ and
$C^{++}$, revealed in the AC model, offered a candidate for dark
matter in the form of elusive (AC)-atoms. In the charge symmetric
case, when primordial concentrations of $A^{--}$ and $C^{++}$ are
equal, a significant fraction of relic $C^{++}$, which is not bound
in (AC)-atoms, is left in the Universe and the suppression of this
fraction in terrestrial matter involves new long range interaction
between A and C, making them to recombine in (AC)-atoms inside dense
matter bodies.

In the asymmetric case, corresponding to excess of -2 charge
species, as it was assumed for $(\bar U \bar U \bar U)^{--}$ in the
model of stable $U$-quark of a 4th generation, their positively
charged partners effectively annihilate in the early Universe. The
dark matter is in the form of nuclear interacting O-helium -
atom-like bound states of -2 charged particles and primordial
helium, formed as soon as $He$ is produced in the Standard Big Bang
Nucleosynthesis (SBBN). Such an asymmetric case was realized in
\cite{KK} in the framework of WTC, where it was possible to find a
relationship between the excess of negatively charged
anti-techni-baryons $(\bar U \bar U )^{--}$ and/or technileptons
$\zeta^{--}$ and the baryon asymmetry of the Universe.

It turned out that the necessary condition for these scenarios,
avoiding anomalous isotopes overproduction, is absence of stable
particles with charge -1, so that stable negatively charged
particles $X^{--}$ should only have charge -2.  After it is formed
in SBBN, $^4He$ screens the $X^{--}$ charged particles in composite
$(^4He^{++}X^{--})$ {\it O-helium} ``atoms''
 \cite{I}. For different models of $X^{--}$ they are also
called ANO-helium \cite{lom,Khlopov:2006dk}, Ole-helium
\cite{FKS,Khlopov:2006dk} or techni-O-helium \cite{KK}. We'll call
them all O-helium ($OHe$) in our further discussion.

In all these forms of O-helium $X^{--}$ behave either as leptons or
as specific "hadrons" with strongly suppressed hadronic interaction.
Therefore O-helium interaction with matter is determined by nuclear
interaction of $He$. These neutral primordial nuclear interacting
objects contribute the modern dark matter density and play the role
of a nontrivial form of strongly interacting dark matter
\cite{Starkman,McGuire:2001qj}. The active influence of this type of
dark matter on nuclear transformations seems to be incompatible with
the expected dark matter properties. However, it turns out that the
considered scenario is not easily ruled out \cite{FKS,I,KK} and
challenges the experimental search for various forms of O-helium and
its charged constituents. O-helium scenario might provide
explanation for the observed excess of positrons in the galactic
bulge and for the controversy between positive results of dark
matter searches in DAMA/NaI (see for review \cite{Bernabei:2003za})
and DAMA/Libra \cite{Bernabei:2008yi} experiments and negative
results of other experimental groups.

\section{O-helium Universe}

Following \cite{I,lom,Khlopov:2006dk,KK} consider charge asymmetric
case, when excess of $X^{--}$ provides effective suppression of
positively charged species.

In the period $100\s \le t \le 300\s$  at $100 \keV\ge T \ge T_o=
I_{o}/27 \approx 60 \keV$, $^4He$ has already been formed in the
SBBN and virtually all free $X^{--}$ are trapped by $^4He$ in
O-helium ``atoms" $(^4He^{++} X^{--})$. Here the O-helium ionization
potential is\footnote{The account for charge distribution in $He$
nucleus leads to smaller value $I_o \approx 1.3 \MeV$
\cite{Pospelov}.} $I_{o} = Z_{x}^2 Z_{He}^2 \alpha^2 m_{He}/2
\approx 1.6 \MeV,$ where $\alpha$ is the fine structure
constant,$Z_{He}= 2$ and $Z_{x}= 2$ stands for the absolute value of
electric charge of $X^{--}$.  The size of these ``atoms" is
\cite{I,FKS} \beq R_{o} \sim 1/(Z_{x} Z_{He}\alpha m_{He}) \approx 2
\cdot 10^{-13} \cm \label{REHe} \eeq

O-helium, being an $\alpha$-particle with screened electric charge,
can catalyze nuclear transformations, which can influence primordial
light element abundance and cause primordial heavy element
formation. These effects need a special detailed and complicated
study. The arguments of \cite{I,FKS,KK} indicate that this model
does not lead to immediate contradictions with the observational
data.

Due to nuclear interactions of its helium constituent with nuclei in
the cosmic plasma, the O-helium gas is in thermal equilibrium with
plasma and radiation on the Radiation Dominance (RD) stage, while
the energy and momentum transfer from plasma is effective. The
radiation pressure acting on the plasma is then transferred to
density fluctuations of the O-helium gas and transforms them in
acoustic waves at scales up to the size of the horizon.

At temperature $T < T_{od} \approx 200 S^{2/3}_3\eV$ the energy and
momentum transfer from baryons to O-helium is not effective
\cite{I,KK} because $n_B \sv (m_p/m_o) t < 1$, where $m_o$ is the
mass of the $OHe$ atom and $S_3= m_o/(1 \TeV)$. Here \beq \sigma
\approx \sigma_{o} \sim \pi R_{o}^2 \approx
10^{-25}\cm^2\label{sigOHe}, \eeq and $v = \sqrt{2T/m_p}$ is the
baryon thermal velocity. Then O-helium gas decouples from plasma. It
starts to dominate in the Universe after $t \sim 10^{12}\s$  at $T
\le T_{RM} \approx 1 \eV$ and O-helium ``atoms" play the main
dynamical role in the development of gravitational instability,
triggering the large scale structure formation. The composite nature
of O-helium determines the specifics of the corresponding dark
matter scenario.

At $T > T_{RM}$ the total mass of the $OHe$ gas with density $\rho_d
= (T_{RM}/T) \rho_{tot} $ is equal to
$$M=\frac{4 \pi}{3} \rho_d t^3 = \frac{4 \pi}{3} \frac{T_{RM}}{T} m_{Pl}
(\frac{m_{Pl}}{T})^2$$ within the cosmological horizon $l_h=t$. In
the period of decoupling $T = T_{od}$, this mass  depends strongly
on the O-helium mass $S_3$ and is given by \cite{KK}\beq M_{od} =
\frac{T_{RM}}{T_{od}} m_{Pl} (\frac{m_{Pl}}{T_{od}})^2 \approx 2
\cdot 10^{44} S^{-2}_3 \g = 10^{11} S^{-2}_3 M_{\odot}, \label{MEPm}
\eeq where $M_{\odot}$ is the solar mass. O-helium is formed only at
$T_{o}$ and its total mass within the cosmological horizon in the
period of its creation is $M_{o}=M_{od}(T_{od}/T_{o})^3 = 10^{37}
\g$.

On the RD stage before decoupling, the Jeans length $\lambda_J$ of
the $OHe$ gas was restricted from below by the propagation of sound
waves in plasma with a relativistic equation of state
$p=\epsilon/3$, being of the order of the cosmological horizon and
equal to $\lambda_J = l_h/\sqrt{3} = t/\sqrt{3}.$ After decoupling
at $T = T_{od}$, it falls down to $\lambda_J \sim v_o t,$ where $v_o
= \sqrt{2T_{od}/m_o}.$ Though after decoupling the Jeans mass in the
$OHe$ gas correspondingly falls down
$$M_J \sim v_o^3 M_{od}\sim 3 \cdot 10^{-14}M_{od},$$ one should
expect a strong suppression of fluctuations on scales $M<M_o$, as
well as adiabatic damping of sound waves in the RD plasma for scales
$M_o<M<M_{od}$. It can provide some suppression of small scale
structure in the considered model for all reasonable masses of
O-helium. The significance of this suppression and its effect on the
structure formation needs a special study in detailed numerical
simulations. In any case, it can not be as strong as the free
streaming suppression in ordinary Warm Dark Matter (WDM) scenarios,
but one can expect that qualitatively we deal with Warmer Than Cold
Dark Matter model.

Being decoupled from baryonic matter, the $OHe$ gas does not follow
the formation of baryonic astrophysical objects (stars, planets,
molecular clouds...) and forms dark matter halos of galaxies. It can
be easily seen that O-helium gas is collisionless for its number
density, saturating galactic dark matter. Taking the average density
of baryonic matter one can also find that the Galaxy as a whole is
transparent for O-helium in spite of its nuclear interaction. Only
individual baryonic objects like stars and planets are opaque for
it.

\section{Detection of O-helium}
The composite nature of O-helium dark matter results in a number of
observable effects.

The nuclear interaction of O-helium with cosmic rays gives rise to
ionization of this bound state in the interstellar gas and to
acceleration of free $X^{--}$ in the Galaxy. Assuming a universal
mechanism of cosmic ray acceleration the anomalous low $Z/A$
component of $-2$ charged $X^{--}$ can be present in cosmic rays and
be within the reach for PAMELA and AMS02 cosmic ray experiments.

Inelastic interaction of O-helium with the matter in the
interstellar space and its de-excitation can give rise to radiation
in the range from few keV to few  MeV. Our first estimations
\cite{FKS,KK} show that the expected signal should be below the
observed gamma background.

However, taking into account that in the galactic bulge with radius
$r_b \sim 1 \kpc$ the number density of O-helium can reach the value
$n_o\approx 3 \cdot 10^{-3}/S_3 \cm^{-3}$, one can estimate the
collision rate of O-helium in this central region: $dN/dt=n_o^2
\sigma v_h 4 \pi r_b^3 /3 \approx 3 \cdot 10^{42}S_3^{-2} \s^{-1}$.
At the velocity of $v_h \sim 3 \cdot 10^7 \cm/\s$ energy transfer in
such collisions is $\Delta E \sim 1 \MeV S_3$. These collisions can
lead to excitation of O-helium. If 2S level is excited, pair
production dominates over two-photon channel in the de-excitation by
$E0$ transition and positron production with the rate $3 \cdot
10^{42}S_3^{-2} \s^{-1}$ is not accompanied by strong gamma signal.
According to \cite{Finkbeiner:2007kk} this rate of positron
production for $S_3 \sim 1$ is sufficient to explain the excess in
positron annihilation line from bulge, measured by INTEGRAL (see
\cite{integral} for review and references). If $OHe$ levels with
nonzero orbital momentum are excited, gamma lines should be observed
from transitions ($ n>m$) $E_{nm}= 1.598 \MeV (1/m^2 -1/n^2)$ (or
from the similar transitions corresponding to the case $I_o = 1.287
\MeV $) at the level $3 \cdot 10^{-4}S_3^{-2}(\cm^2 \s \MeV
ster)^{-1}$.

The evident consequence of the O-helium dark matter is its
inevitable presence in the terrestrial matter, which appears opaque
to O-helium and stores all its in-falling flux.

If the $OHe$ diffusion in matter is determined by elastic
collisions, the in-falling $OHe$ particles are effectively slowed
down after they fall down terrestrial surface. Then they drift,
sinking down towards the center of the Earth with velocity \beq V =
\frac{g}{n \sigma v} \approx 80 S_3 A^{1/2} \cm/\s. \label{dif}\eeq
Here $A \sim 30$ is the average atomic weight in terrestrial surface
matter, $n=2.4 \cdot 10^{24}/A$ is the number of terrestrial atomic
nuclei, $\sigma v$ is the rate of nuclear collisions and $g=980~
\cm/\s^2$.

Near the Earth's surface, the O-helium abundance is determined by
the equilibrium between the in-falling and down-drifting fluxes.
Such neutral $(^4He^{++}X^{--})$ ``atoms" may provide a catalysis of
cold nuclear reactions in ordinary matter (much more effectively
than muon catalysis). This effect needs a special and thorough
investigation. On the other hand, $X^{--}$ capture by nuclei,
heavier than helium, can lead to production of anomalous isotopes,
but the arguments, presented in \cite{I,FKS,KK} indicate that their
abundance should be below the experimental upper limits.

It should be noted that the nuclear cross section of the O-helium
interaction with matter escapes the severe constraints
\cite{McGuire:2001qj} on strongly interacting dark matter particles
(SIMPs) \cite{McGuire:2001qj,Starkman} imposed by the XQC experiment
\cite{XQC}. Therefore, a special strategy of direct O-helium  search
is needed, as it was proposed in \cite{Belotsky:2006fa}.

In underground detectors, $OHe$ ``atoms'' are slowed down to thermal
energies and give rise to energy transfer $\sim 2.5 \cdot 10^{-4}
\eV A/S_3$, far below the threshold for direct dark matter
detection. It makes this form of dark matter insensitive to the
severe CDMS constraints \cite{Akerib:2005kh}. However, $OHe$ induced
nuclear transformations can result in observable effects.

At a depth $L$ below the Earth's surface, the drift timescale is
$t_{dr} \sim L/V$, where $V \sim 400 S_3 \cm/\s$ is given by
Eq.~(\ref{dif}). It means that the change of the incoming flux,
caused by the motion of the Earth along its orbit, should lead at
the depth $L \sim 10^5 \cm/\s$ to the corresponding change in the
equilibrium underground concentration of $OHe$ on the timescale
$t_{dr} \approx 2.5 \cdot 10^2 S_3^{-1}\s$. Such rapid adjustment of
local fraction of $OHe$ provides annual modulations of inelastic
processes inside the bodies of underground dark matter detectors.

One can expect two kinds of inelastic processes in the matter with
nuclei $(A,Z)$, having atomic number $A$ and charge $Z$ \beq
(A,Z)+(HeX) \rightarrow (A+4,Z+2) +X^{--}, \label{EHeAZ} \eeq and
\beq (A,Z)+(HeX) \rightarrow [(A,Z)X^{--}] + He. \label{HeEAZ} \eeq

The first reaction is possible, if the masses of the initial and
final nuclei satisfy the energy condition \beq M(A,Z) + M(4,2) -
I_{o}> M(A+4,Z+2), \label{MEHeAZ} \eeq where $I_{o} = 1.6 \MeV$ is
the binding energy of O-helium and $M(4,2)$ is the mass of the
$^4He$ nucleus. It is more effective for lighter nuclei, while for
heavier nuclei the condition (\ref{MEHeAZ}) is not valid and
reaction (\ref{HeEAZ}) should take place.

In the both types of processes energy release is of the order of
MeV, which seems to have nothing to do with the signals in the DAMA
experiment. However, in the reaction (\ref{HeEAZ}) such energy is
rapidly carried away by $He$ nucleus, while in the remaining
compound state of $[(A,Z)X^{--}]$ the charge of the initial $(A,Z)$
nucleus is reduced by 2 units and the corresponding transformation
of electronic orbits with possible emission of two excessive
electrons should take place. The energy difference between the $K$
orbits of the lowest lying electronic $1s$ level of the initial
nucleus with the charge $Z$ and the respective levels of its
compound system with $X^{--}$ is given by \beq \Delta E
=Z^2\alpha^2m_e/2 -(Z-2)^2\alpha^2m_e/2] \approx Z\alpha^2m_e.
\label{DEHeAZ} \eeq Here we took into account that the energy
difference comes from the change in the nuclear charge with the
initially unchanged structure of electronic shells. It is
interesting that the energy release in such transition for two $1s$
electrons in $^{53}I_{127}$ is about 2 keV, while for
$^{81}Tl_{205}$ it is about 4 keV. Taking into account that the
signal in the DAMA experiment was detected with similar energy of
ionization, this idea deserves more detailed analysis, which might
be useful for interpretation of this experiment. Since the
experimental cuts in the CDMS experiment \cite{Akerib:2005kh},
exclude events of pure ionization, which are not accompanied by
phonon signal, if valid, the proposed mechanism could explain the
difference in the results of DAMA and CDMS and other direct dark
matter searches that imply nuclear recoil measurement, which should
accompany ionization. We have discussed a possibility for such
explanation in the framework of the minimal Walking Technicolor
model in \cite{KK2}.

An inevitable consequence of the proposed explanation is appearance
in the matter of DAMA/NaI or DAMA/Libra detector anomalous
superheavy isotopes of antinomy (Sb with nuclear charge $Z=53-2=51$)
and gold (Au with nuclear charge $Z=81-2=79$), created in the
inelastic process (\ref{HeEAZ}) and having the mass roughly by $m_o$
larger, than ordinary isotopes of these elements. If the atoms of
these anomalous isotopes are not completely ionized, their mobility
is determined by atomic cross sections and becomes about 9 orders of
magnitude smaller, than for O-helium. It provides conservation in
the matter of detector of at least 200 anomalous atoms per 1g,
corresponding to the number of events, observed in DAMA experiment.
Therefore mass-spectroscopic analysis of this matter can provide
additional test for the O-helium nature of DAMA signal.

\section*{Acknowledgments}
I am grateful to P.Belli, K.Belotsky, J.Filippini, C.Kouvaris, F.
Lebrun, A.Mayorov, P.Picozza, V.Rubakov, E.Soldatov and D.Spergel
for discussions and to S.Katsanevas for support.

\end{document}